%% file: HZZnote.tex
\documentclass[11pt,a4paper]{scrartcl}

\usepackage{CLICdp}

\usepackage{CLICdp_definitions}
\usepackage{pst-pdf}
\usepackage[symbol]{footmisc}
\usepackage{feynmf}
\usepackage{textpos}
\usepackage{lineno}

\title{Measurement of ${\sigma(H\nu_e\bar{\nu_e})\times BR(H\rightarrow ZZ^\ast)}$ and Higgs production in $ZZ$ fusion at a 1.4 TeV CLIC collider}

\clicdpconf{2015}{004}  

\date{\today}

\addauthor{G.~Milutinovi\'{c}-Dumbelovi\'{c}\thanks{gordanamd@vinca.rs}}{\institute{1}}
\addauthor{I.~Bo\v{z}ovi\'{c}-Jelisav\u{c}i\'{c}}{\institute{1}}
\addauthor{A.~Robson}{\institute{2}}
\addauthor{P.~Roloff}{\institute{3}}

\addinstitute{1}{Vinca Institute of Nuclear Sciences, University of Belgrade, Serbia}
\addinstitute{2}{University of Glasgow, UK}
\addinstitute{3}{CERN, Switzerland}

\onbehalfof{the CLICdp collaboration}

\abstract{This paper presents the potential measurement at 1.4 TeV CLIC of the
cross-section (times branching ratio) of the Higgs production via $WW$
fusion with the Higgs subsequently decaying in $ZZ^\ast$, ${\sigma(H\nu_e\bar{\nu_e})\times BR(H\rightarrow ZZ^\ast)}$, 
and of the Higgs production via $ZZ$ fusion with the Higgs
subsequently decaying in $b\bar{b}$, ${\sigma(He{^+}e{^-})\times BR(H\rightarrow  b\bar{b})}$.
For the $H\rightarrow ZZ^\ast$ decay the hadronic final state, ${ZZ^\ast\rightarrow  q\bar{q}q\bar{q}}$, and the semi-leptonic final state, ${ZZ^\ast\rightarrow  q\bar{q}l^+l^-}$, are considered.
The results show that ${\sigma(H\nu_e\bar{\nu_e})\times BR(H\rightarrow ZZ^\ast)}$ can be measured with a
precision of 18.3\% and 6\% for the hadronic and semi-leptonic channel, respectively.
${\sigma(He{^+}e{^-})\times BR(H\rightarrow  b\bar{b})}$ can be measured with a precision of 1.7\%.
This measurement also contributes to the determination of the Higgs
coupling to the $Z$ boson, $g_{H_{ZZ}}$.}

\titlecomment{Talk presented at the International Workshop on Future Linear Colliders (LCWS14), Belgrade, Serbia, 6--10 October 2014.}

\begin{document}
\titlepage

\section{Introduction} 
\input{introduction.tex}

\section{Simulation and Reconstruction}
\input{simulation.tex}

\section{The CLIC\_ILD detector model}
\label{detector}
\input{detector.tex}

\section{Measurement of ${\sigma(H\nu\bar{\nu})\times BR(H\rightarrow ZZ^\ast)}$}
\subsection{Event samples}
\label{event_samples}
\input{samples.tex}

\subsection{Analysis strategy}
\label{analysis_strategy}
\input{analysis_strategy.tex}

\subsubsection{Lepton identification}
\label{lepton_identification}
\input{lepton_identification.tex}

\subsubsection{Preselection}
\label{preselection}
\input{preselection.tex}

\subsubsection{MVA event selection}
\label{MVA}
\input{MVA.tex}

\section{Higgs production in ZZ fusion and ${\sigma(He{^+}e{^-})\times BR(H\rightarrow  b\bar{b})}$ measurement }

\subsection{Event samples}
\label{event_samplesbb}
\input{samplesbb.tex}

\subsection{Analysis strategy}
\label{analysisstrategybb}
\input{analysis_strategybb.tex}

\subsubsection{Preselection}
\label{preselectionbb}
\input{preselectionbb.tex}

\subsubsection{MVA}
\label{MVAbb}
\input{MVAbb.tex}

\section{Results}
\label{results}
\input{results.tex}

\printbibliography[title=References]

\end{document}

%% file: introduction.tex
The Compact Linear Collider (CLIC) is a proposed high-luminosity linear $e^+e^-$ collider planned to be implemented in stages with centre-of-mass energy, ${\sqrt{s}}$, 
of 350 GeV (or more), 1.4 TeV and 3 TeV. One of the main aims of CLIC would be the high precision measurement of the Higgs boson properties \cite{CLIC_PhysDet_CDR, CLIC_snowmass13}.

This paper focuses on the measurement contributing to the determination of the Higgs coupling to the $Z$ boson, $g_{H_{ZZ}}$, at 1.4 TeV CLIC.

At ${\sqrt{s} =1.4}$ TeV, the dominant Higgs production process is the $WW$ fusion, with $\sim$ 370000 expected events in 1.5 ab$^{-1}$ of data. This would lead to the measurement of the
relative coupling of the Higgs boson to the $W$ and $Z$ bosons at the percent level, providing a strong test of the Standard Model prediction for $g_{H_{\PW\PW}}$/$g_{H_{\PZ\PZ}}=cos^2 \theta_W$,
where $\theta_W$ is the Weinberg angle.

The subleading Higgs production process is the $ZZ$ fusion that, with $\sim$ 10\% the cross section of the $WW$ fusion, would give $\sim$ 37000 expected events in 1.5 ab$^{-1}$ of data
and it would provide access to complementary Higgs boson coupling, $g_{H_{ZZ}}$, at a percent level.

In this paper, we discuss the measurement of the ${\sigma(H\nu_e\bar{\nu_e})\times BR(H\rightarrow ZZ^\ast)}$ where Higgs is produced via $WW$ fusion, and the fully hadronic 
${ZZ^\ast\rightarrow  q\bar{q}q\bar{q}}$, and semi-leptonic ${ZZ^\ast\rightarrow  q\bar{q}l^+l^-}$ final states are considered. For the Higgs production in $ZZ$ fusion, the $H\rightarrow  b\bar{b}$ decay is analysed.

Both analyses are performed using the CLIC\_ILD detector concept \cite{LCDnote_CLICILDCDRgeo}, assuming a total integrated luminosity of 1.5 ab$^{-1}$
and unpolarised beams.

%% file: simulation.tex
The Higgs production through $WW$ and $ZZ$ fusion is generated in Whizard 1.95 \cite{Kilian:2007gr}, where a Higgs mass of 126 GeV is assumed. 
The background events are also generated in Whizard, using Pythia 6.4 \cite{Sjostrand2006} to simulate the hadronisation and fragmentation processes. 
The CLIC luminosity spectrum and the beam-induced processes are obtained from GuineaPig 1.4.4 \cite{Schulte:1999tx}. All events are simulated with unpolarised beams.

The interactions with the detector are simulated using the CLIC\_ILD detector model within the Mokka simulation package \cite{Mora2002}, based on GEANT4 \cite{Agostinelli2003}. 

Events are reconstructed using the particle-flow technique, implemented in the Pandora algorithm (PFA) \cite{thomson:pandora, Marshall2013153}. 
The $k_t$ algorithm \cite{S.Catani}, implemented in FastJet \cite{FastJet}, is used in the
exclusive mode to cluster the jets of each event. The LCFIPlus package \cite{LCFIPlus} is used for the identification of charm and beauty jets. 
The IsolatedLeptonFinder Marlin processor \cite{IsolatedLeptonFinder} is used to identify leptons. The TMVA package \cite{TMVA:2010} is used for the multivariate classification of signal and background events using 
their kinematic properties. 

The simulation, reconstruction and analysis are done with ILCDIRAC \cite{ILCDIRAC}.

%% file: detector.tex
The CLIC\_ILD detector is based on the ILD detector concept \cite{ildloi:2009} for ILC, 
modified according to specific experimental conditions at CLIC \cite{CLIC_PhysDet_CDR}. 
The main tracking device of CLIC\_ILD is a Time Projection Chamber providing a point resolution in the $r\phi$ plane better than 100 ${\upmu}$m. 
The precision physics at CLIC require a vertex-detector system with excellent flavour-tagging capabilities through the measurement of displaced vertices.
The vertex detector is based on ultra-thin hybrid pixel sensors technology, and uses power-pulsing and air-flow cooling to minimise the material budget.
The CLIC\_ILD detector concept is based on fine-grained electromagnetic and hadronic calorimeters (ECAL and HCAL), optimised for particle-flow techniques. 
Both calorimeters are within a 4 T solenoidal magnetic field.

%% file: samples.tex
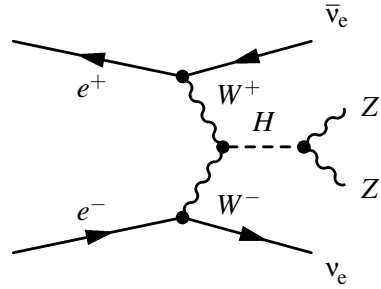
\begin{figure}
\centering
\begin{fmffile}{wwfusion}
\begin{center}
\begin{fmfgraph*}(140,80) 
\fmfleft{i1,i2} \fmfright{o1,o2,o3,o4}
\fmf{fermion,label=$e^-$}{i1,v1} \fmf{fermion,label=$e^+$,label.side=left}{v3,i2}
\fmf{fermion}{v1,o1} \fmf{fermion}{o4,v3}
\fmf{boson,label=$W^-$}{v1,v2} \fmf{boson,label=$W^+$,label.side=left}{v3,v2} 
\fmf{dashes,label=$H$}{v2,v4}
\fmf{boson}{v4,o2} \fmf{boson}{o3,v4} 
\fmfdot{v1,v2,v3,v4}
\fmflabel{$Z$}{o2} \fmflabel{$Z$}{o3}
\fmflabel{$\PGne$}{o1} \fmflabel{$\PAGne$}{o4}
\end{fmfgraph*}
\end{center}
\end{fmffile}
\caption{\label{fig-ww} Feynman diagram of the Higgs production in $WW$ fusion and the subsequent Higgs boson decay to a pair of $Z$ bosons.}
\end{figure}

The $WW$ fusion process has the largest cross section for Higgs production at 1.4 TeV CLIC.
The cross section for $e{^+}e{^-}\rightarrow H\nu\bar{\nu}$ is 244 fb. The branching fraction for the $H\rightarrow ZZ^\ast$ decay is 2.89\% \cite{Dittmaier:2012vm}. 
The Feynman diagram for the process $e^-e^+\rightarrow H\nu\bar{\nu}\rightarrow \PZ\PZ^\ast \nu\bar{\nu}$ is shown in Figure \ref{fig-ww}.

In this analysis, two final states of the ${ZZ^\ast}$ decays are studied: the fully hadronic final state, $ZZ^\ast\rightarrow q\bar{q}q\bar{q}$, with a
branching ratio of 49\%, resulting in an effective cross section of 3.45 fb and the semi-leptonic final state, $ZZ^\ast\rightarrow q\bar{q}l^+l^-$, with a branching ratio of 10\% 
and an effective cross-section of 0.72 fb.

In Table \ref{tableprocess}, the full list of the signal and background processes is given with the corresponding cross-sections. 

The main background, characterised by the same final state as the fully hadronic signal final state, is given by the $e^-e^+\rightarrow H\nu_e\bar{\nu_e}\rightarrow \PW\PW^\ast \nu_e\bar{\nu_e}\rightarrow q\bar{q}q\bar{q}\nu_e\bar{\nu_e}$ process.
Other important backround processes due to their large cross section are ${e^\pm}\gamma\rightarrow q\bar{q}q\bar{q}\nu$,
$\gamma\gamma\rightarrow q\bar{q}q\bar{q}$ and ${e^\pm}\gamma\rightarrow q\bar{q}q\bar{q}e$. They can be substantially reduced by requiring high-$\text p_{\text T}$ jets.
Other minor background processes can be discriminated from signal events using a MVA event classifier.

\begin{table} 
\centering
\caption{\label{tableprocess} List of considered processes with the corresponding cross-sections. Background processes marked by \dag \/, where jets and leptons are 
not produced by a Higgs boson, are generated with a cut on the Higgs mass set to $m_{H}$=12 TeV.
Because of the large cross section of the processes marked by *, generator level cuts are introduced to limit the number of events that needed to be simulated 
and reconstructed. The invariant mass of the two jets, $m_{q\bar{q}}$, is required to be greater than 50 GeV and invariant mass of four jets, $m_{q\bar{q}q\bar{q}}$ is
also set to be greater than 50 GeV. The cross sections for all processes with photons in the initial state include both processes with the Beamstrahlung 
and with Effective Photon Approximation (EPA) photons. 
Cross sections for the processes ${e^\pm}\gamma\rightarrow q\bar{q}\nu$, ${e^\pm}\gamma\rightarrow q\bar{q}q\bar{q}\nu$, ${e^\pm}\gamma\rightarrow q\bar{q}q\bar{q}e$,
and ${e^\pm}\gamma\rightarrow q\bar{q}e$ represent the sum of processes with the electron and the positron in the initial state.}
\begin{tabular*}{\columnwidth}{@{\extracolsep{\fill}}ll@{}}
\hline
\multicolumn{1}{@{}l}{Signal process}  & $\sigma(fb)$ \\
\hline
$e^-e^+\rightarrow H\nu\bar{\nu}, H\rightarrow \PZ\PZ^\ast, \PZ\PZ^\ast\rightarrow q\bar{q}q\bar{q}$              & 3.45         \\
$e^-e^+\rightarrow H\nu\bar{\nu}, H\rightarrow \PZ\PZ^\ast, \PZ\PZ^\ast\rightarrow q\bar{q}l^+l^-$              & 0.72         \\
\hline     
\multicolumn{1}{@{}l}{Common background}  & $\sigma(fb)$ \\  
\hline 
$e^-e^+\rightarrow q\bar{q}\nu_{e}\bar{\nu_{e}}$                      & 788        \\
$e^-e^+\rightarrow q\bar{q}q\bar{q}\nu_{e}\bar{\nu_{e}}$                 & 24.7         \\
$e^-e^+\rightarrow H\nu\bar{\nu}, H\rightarrow \PW\PW, \PW\PW\rightarrow q\bar{q}q\bar{q}$   & 27.6         \\
$e^-e^+\rightarrow H\nu\bar{\nu}, H\rightarrow b\bar{b}$                     & 136.94        \\
$e^-e^+\rightarrow q\bar{q}$                      & 4009.5        \\
$e^-e^+\rightarrow q\bar{q}q\bar{q}$                      & 1245.1${^\dag}$        \\
$e^-e^+\rightarrow q\bar{q}q\bar{q}l^+l^-$                      & 71.7       \\
$e^-e^+\rightarrow q\bar{q}q\bar{q}l\nu$                      & 115.3      \\
$e^-e^+\rightarrow H\nu\bar{\nu}, H\rightarrow \PZ\PZ^\ast, \PZ\PZ^\ast\rightarrow l^+l^-l^+l^-$      & 0.08       \\
\hline
\multicolumn{1}{@{}l}{$e^-e^+\rightarrow H\nu\bar{\nu}, H\rightarrow \PZ\PZ^\ast, \PZ\PZ^\ast\rightarrow q\bar{q}q\bar{q}$ specific background}  & $\sigma(fb)$ \\
\hline
${e^\pm}\gamma\rightarrow q\bar{q}q\bar{q}\nu$                       & 338.5${^\dag}$        \\
$\gamma\gamma\rightarrow q\bar{q}q\bar{q}$                  & 30212${^*\dag}$      \\
${e^\pm}\gamma\rightarrow q\bar{q}q\bar{q}e$                       & 2891       \\
$e^-e^+\rightarrow H\nu\bar{\nu}, H\rightarrow \PZ\PZ^\ast, \PZ\PZ^\ast\rightarrow q\bar{q}l^+l^-$              & 0.72         \\
\hline
\multicolumn{1}{@{}l}{$e^-e^+\rightarrow H\nu\bar{\nu}, H\rightarrow \PZ\PZ^\ast, \PZ\PZ^\ast\rightarrow q\bar{q}l^+l^-$ specific background}  & $\sigma(fb)$ \\
\hline
$e^-e^+\rightarrow q\bar{q}l^+l^-$                              & 2725.8${^\dag}$       \\
${e^\pm}\gamma\rightarrow q\bar{q}\nu$                             & 37125.3 ${^*\dag}$       \\
${e^\pm}\gamma\rightarrow q\bar{q}e$                             & 63838.8${^*\dag}$       \\
$\gamma\gamma\rightarrow q\bar{q}$                             & 112038.6 ${^*\dag}$       \\
$e^-e^+\rightarrow H\nu\bar{\nu}, H\rightarrow \PZ\PZ^\ast, \PZ\PZ^\ast\rightarrow q\bar{q}q\bar{q}$              & 3.45         \\
\hline
\end{tabular*}
\end{table}

%% file: analysis_strategy.tex
The following section describes the physics object identification and the selection requirements applied in the analysis.

For the semi-leptonic final state, the first step of the physics object identification is searching for isolated leptons (electrons or muons).
Exactly two leptons are required, otherwise the event is rejected. Then, all particles in the event not identified as leptons are clustered 
by the $k_t$ algorithm into two jets with $R$ = 1.0.

Instead, for the hadronic final state, the event is directly clustered by the $k_t$ algorithm into four jets with $R$ = 1.0.

Next, for both final states, flavour-tagging is performed and a preselection based on kinematics variables is applied. Finally, a MVA event selection based on the BDT
classifier is performed to obtain the final results.

%% file: lepton_identification.tex
The first step of the semi-leptonic analysis is to identify and reconstruct leptons from $ZZ^\ast$ decays. This section
describes how leptons are distinguished from all other reconstructed particles in the event.
Isolated leptons are identified using a combination of track energy, calorimeter energy and impact parameter. 

Muons and electrons are required to have a track energy of at least
7 GeV. The impact parameter of a track describes the perpendicular distance between the track and primary vertex (PV), at the track\textquoteright s 
point of closest approach to the PV. It can be decomposed into longitudinal ($Z_0$) and radial ($d_0$)
components, which combine to give the impact parameter in 3 dimension ($R_0$) 

\begin{equation}\label{eq-R0}
R_0 = \sqrt{{Z_0}^2+{d_0}^2} .
\end{equation}

In the analysis, an impact parameter smaller than 0.02 mm is required.

The ratio (RCAL) of the energy deposits in the ECAL and in the HCAL

\begin{equation}
RCAL =E_{ECAL}/(E_{ECAL}+E_{HCAL})
\label{eq-RCAL}
\end{equation}

is a powerful discriminating variable.

Since electrons are mostly contained within the ECAL, they have a peak at RCAL=1. Muons deposit a minimal amount of ionisation energy throughout the calorimeters, and have a peak at
RCAL=0.2. Therefore, in order to reduce the miss-identification lepton contribution, RCAL is required to be either larger than 0.9 or in the range 0.025-0.300. 

Applying these selection criteria, 74\% of electron and muon pairs from Z decays are correctly identified.

%% file: preselection.tex
Leptons and jets are paired to give the $Z$ bosons contribution. Since $m_{H}$ < 2$m_{Z}$, one pair is required to have a mass consistent with $m_{Z}$ (on-shell $Z$
boson), while the second pair is required to form the off-shell $Z$ boson. The preselection cuts require: 
\begin{itemize}
\item on-shell Z boson: 45 GeV < $m_{Z}$ < 110 GeV, 
\item off-shell Z boson: $m_{Z^\ast}$ < 65 GeV,
\item Higgs invariant mass: 90 GeV < $m_{H}$ < 165 GeV,
\item the distance value between the two closest jets (jet transition cuts): $-\log y_{34}$ < 3.5, $-\log y_{23}$ < 3.0, 
\item visible energy: 100 GeV < $E_{vis}$ < 600 GeV,
\item missing transverse momentum $p_{\mathrm T}^{\mathrm{miss}}$ > 80 GeV, 
\item In order to reject $H\rightarrow b\bar{b}$ decays, the event is forced into a two-jet topology and the
flavour-tag is applied to the two jets. Events where one or both jets have a b-tag probability, $P(b)^{(jet)}$, greater than 0.95 are rejected.
\end{itemize}

The jet transition values, $-\log y_{34}$ and  $-\log y_{23}$, are used in the preselection to discriminate signal from lower jet multiplicity backgrounds.

In addition, for the semi-leptonic final state exactly two isolated leptons are required.

The signal preselection efficiency is 30.2\% for the fully-hadronic final state and 74\% for the semi-leptonic final state. The preselection
efficiency  in the fully-hadronic final state is relatively low due to  $p_{\mathrm T}^{\mathrm{miss}}$> 80 GeV cut in order to suppress the ${e^\pm}\gamma\rightarrow q\bar{q}q\bar{q}\nu$ and ${e^\pm}\gamma\rightarrow q\bar{q}q\bar{q}e$ processes.

%% file: MVA.tex
After the preselection, a MVA event selection based on the BDT classifier is applied in the analysis.

For the fully hadronic final state, the following 11 sensitive observables are used for the classification of the events:
$m_{Z}$, $m_{Z^\ast}$, $m_{H}$, $-\log y_{34}$, $-\log y_{23}$, $E_{vis}$, $p_{T}$, $P(b)^{jet_1}$, $P(b)^{jet_2}$, $P(c)^{jet_1}$ and $P(c)^{jet_2}$; while for the semi-leptonic final state
the following 17 observables are used: $m_{Z}$, $m_{Z^\ast}$, $m_{l^+l^-}$, $m_{q\bar{q}}$, $m_{H}$, $E_{vis}$, $-\log y_{34}$, $-\log y_{23}$, $-\log y_{12}$, $P(b)^{jet_1}$, $P(b)^{jet_2}$, $P(c)^{jet_1}$, 
$P(c)^{jet_2}$, $p_{T}$, $\theta_{H}$, $E_{vis}-E_{H}$ and $N_{PFO}$. These variables are defined as follows: $N_{PFO}$ is the number of all particle-flow objects in one event, $m_{l^+l^-}$ is the invariant mass of the two 
selected leptons, $m_{q\bar{q}}$ is the invariant mass of the two selected jets, $m_{Z}$ is the invariant mass of the on-shell $Z$ boson, $m_{Z^\ast}$ is the the invariant mass of the off-shell $Z$ boson,
$m_{H}$ is the invariant mass of the Higgs candidate, $E_{vis}$ is the visible energy of the event,  $E_{vis}-E_{H}$ is
the difference between the visible energy in the event and the Higgs visible energy,
$-\log y_{34}$,$-\log y_{23}$ and $-\log y_{12}$ are transition variables, $P(b)^{jet_1}$ and $P(b)^{jet_2}$ are b-tag probability of the jets, $P(c)^{jet_1}$ and 
$P(c)^{jet_2}$ are c-tag probability of the jets, $p_{T}$ is missing transverse momentum and $\theta_{H}$ is polar angle of the Higgs candidate.

The BDT is trained on  $H\rightarrow b\bar{b}$, $e^-e^+\rightarrow q\bar{q}\nu_{e}\bar{\nu_{e}}$ and ${e^\pm}\gamma\rightarrow q\bar{q}q\bar{q}\nu$ background samples and the signal sample
for the fully hadronic final state, while it is trained on all the background sample and signal for the semi-leptonic final state.

\begin{figure*}
\centering
\includegraphics[width=12cm, clip, trim = 0.5cm 5cm 0.5cm 3cm]{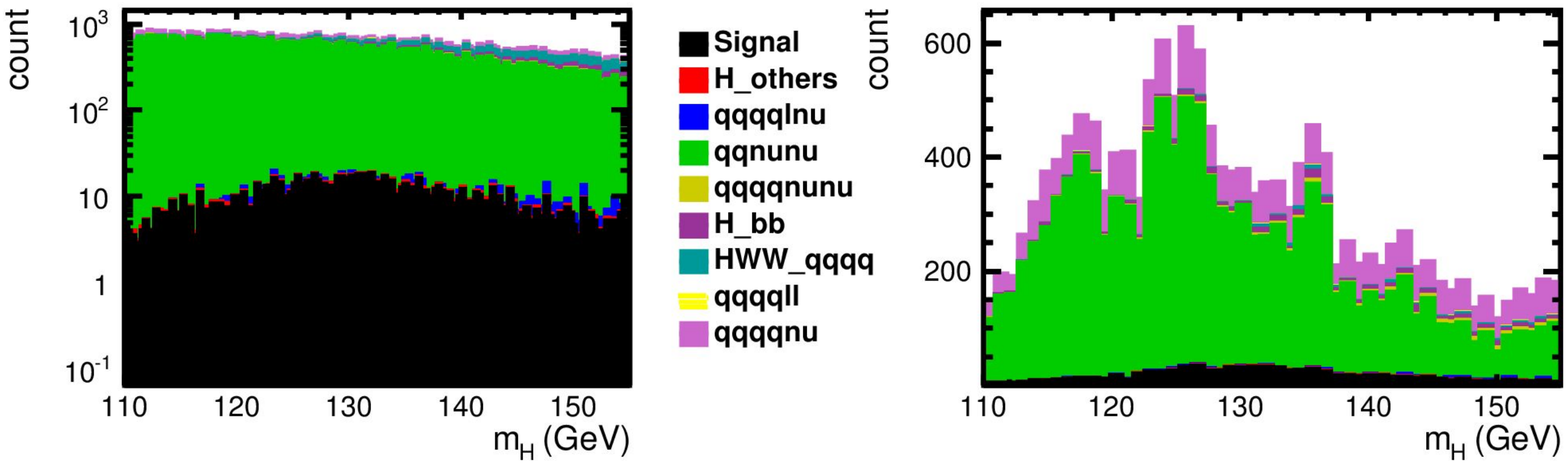}
\begin{textblock}{0.2}(3.2,-3.5)
\bf CLICdp
\end{textblock}
\begin{textblock}{0.2}(12,-3.5)
\bf CLICdp
\end{textblock}
\caption{\label{fig-stackplot} The Higgs invariant mass distributions after preselection (left) and after MVA selection (right) for the fully-hadronic final state.}
\end{figure*}

\begin{figure*}
\centering
\includegraphics[width=12cm, clip, trim = 0.5cm 5cm 0.5cm 3cm]{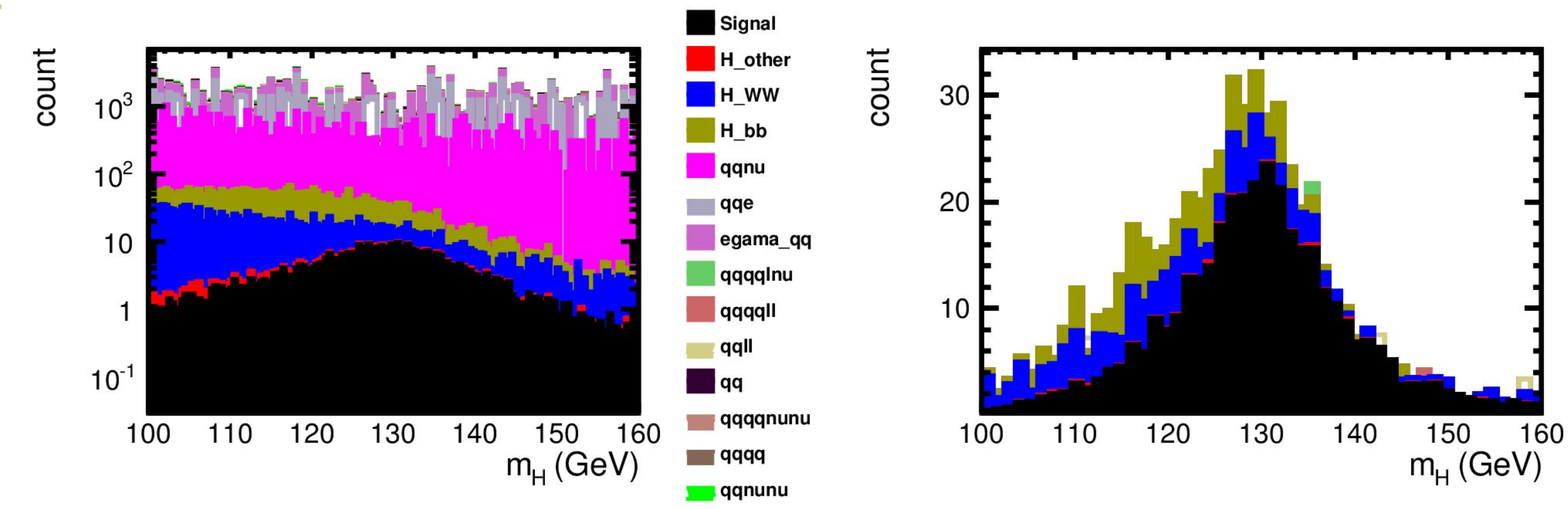}
\begin{textblock}{0.2}(3.2,-3.5)
\bf CLICdp
\end{textblock}
\begin{textblock}{0.2}(12,-3.5)
\bf CLICdp
\end{textblock}
\caption{\label{fig-stackqqll} The Higgs invariant mass distributions after preselection (left) and after MVA selection (right) for the semi-leptonic final state.}
\end{figure*}

In both final states, the BDT cut maximising the significance is chosen, giving an overall efficiency of 18\% and 33\%, for the fully-hadronic and semi-leptonic final 
states, respectively.
Figure \ref{fig-stackplot} (left) includes all events that pass the preselection, while Figure \ref{fig-stackplot} (right) shows all events passing the BDT selection for the 
fully-hadronic final state. The same is shown for the semi-leptonic final state in Figure \ref{fig-stackqqll}. 
All samples are normalised to the integrated luminosity of 1.5 ab$^{-1}$. 

The final results and the statistical uncertainty for the ${\sigma(H\nu_e\bar{\nu_e})\times BR(H\rightarrow ZZ^\ast)}$ measurement are reported in Section \ref{results}.

%% file: samplesbb.tex
The $ZZ$ fusion process has a 10 times smaller cross section for Higgs production at 1.4 TeV CLIC than the $WW$ fusion process. 
The characteristic signature of the $ZZ$ fusion process is two
scattered beam electrons with a large pseudorapidity separation, plus the Higgs boson decay products. 
The cross section for $e{^+}e{^-}\rightarrow He^-e^+$ is 25 fb. 
In this analysis, the scattered beam electrons are required to be fully reconstructed, and the final state
$H\rightarrow b\bar{b}$ is considered. The branching fraction for the $H\rightarrow b\bar{b}$ 
decay is 56.1\%. The corresponding cross-section for the signal is 13.74 fb. 

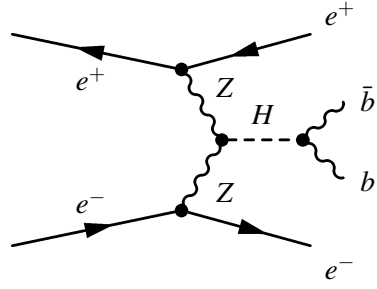
\begin{figure}
\centering
\begin{fmffile}{zzfusion}
\begin{center}
\begin{fmfgraph*}(140,80) 
\fmfleft{i1,i2} \fmfright{o1,o2,o3,o4}
\fmf{fermion,label=$e^-$}{i1,v1} \fmf{fermion,label=$e^+$,label.side=left}{v3,i2}
\fmf{fermion}{v1,o1} \fmf{fermion}{o4,v3}
\fmf{boson,label=$Z$}{v1,v2} \fmf{boson,label=$Z$,label.side=left}{v3,v2} 
\fmf{dashes,label=$H$}{v2,v4}
\fmf{boson}{v4,o2} \fmf{boson}{o3,v4} 
\fmfdot{v1,v2,v3,v4}
\fmflabel{$b$}{o2} \fmflabel{$\bar{b}$}{o3}
\fmflabel{$e^-$}{o1} \fmflabel{$e^+$}{o4}
\end{fmfgraph*}
\end{center}
\end{fmffile}
\caption{\label{fig-zz} Feynman diagram of the Higgs production in $ZZ$ fusion and the subsequent Higgs boson decay to a pair of $b$ quarks.}
\end{figure}

In Figure \ref{fig-zz} the Feynman diagram for 
$e{^+}e{^-}\rightarrow He^-e^+ \rightarrow b\bar{b}e^-e^+$ is shown. In Table \ref{tableprocess2} a full list of the signal and background processes is given with the corresponding cross-sections. 

The main backgound is $e^-e^+\rightarrow q\bar{q}l^+l^-$. It can be reduced requiring b-tagged jets. Other background processes give very small contributions after the preselection.

\begin{table} 
\centering
\caption{\label{tableprocess2} List of considered processes with the corresponding cross-sections.}
\begin{tabular*}{\columnwidth}{@{\extracolsep{\fill}}ll@{}}
\hline
\multicolumn{1}{@{}l}{Process}  & $\sigma(fb)$ \\
\hline
$e^-e^+\rightarrow H e^-e^+, H\rightarrow b\bar{b}$              & 13.74         \\
$e^-e^+\rightarrow  q\bar{q}l^+l^-$                      & 2727        \\
$e^-e^+\rightarrow q\bar{q}q\bar{q}$                      & 1328        \\
$e^-e^+\rightarrow q\bar{q}q\bar{q}l^+l^-$                      & 71.7       \\
$e^-e^+\rightarrow t\bar{t}$                      & 135.8      \\
$\gamma\gamma\rightarrow q\bar{q}e^-e^+$                      & 6       \\
\hline
\end{tabular*}
\end{table}

%% file: analysis_strategybb.tex
This section describes the analysis method for the reconstruction of Higgs candidates produced in $ZZ$ fusion undergoing subsequent decay into $b\bar{b}$.
Events are clustered into a 4-jet topology using a $k_t$ exclusive clustering algorithm with $R$ = 1.0. For a well-reconstructed signal event, 
two of the resulting \textquoteleft jets\textquoteright  are expected to be reconstructed electrons, and the remaining two jets are from
the Higgs decay to $b\bar{b}$.
The discrimination between signal and background events is based on pre-selection
cuts and a multivariate likelihood analysis.

%% file: preselectionbb.tex
\begin{figure*}
\centering
\includegraphics[width=6cm, clip, trim = 0.05cm 8cm 8cm 0.5cm]{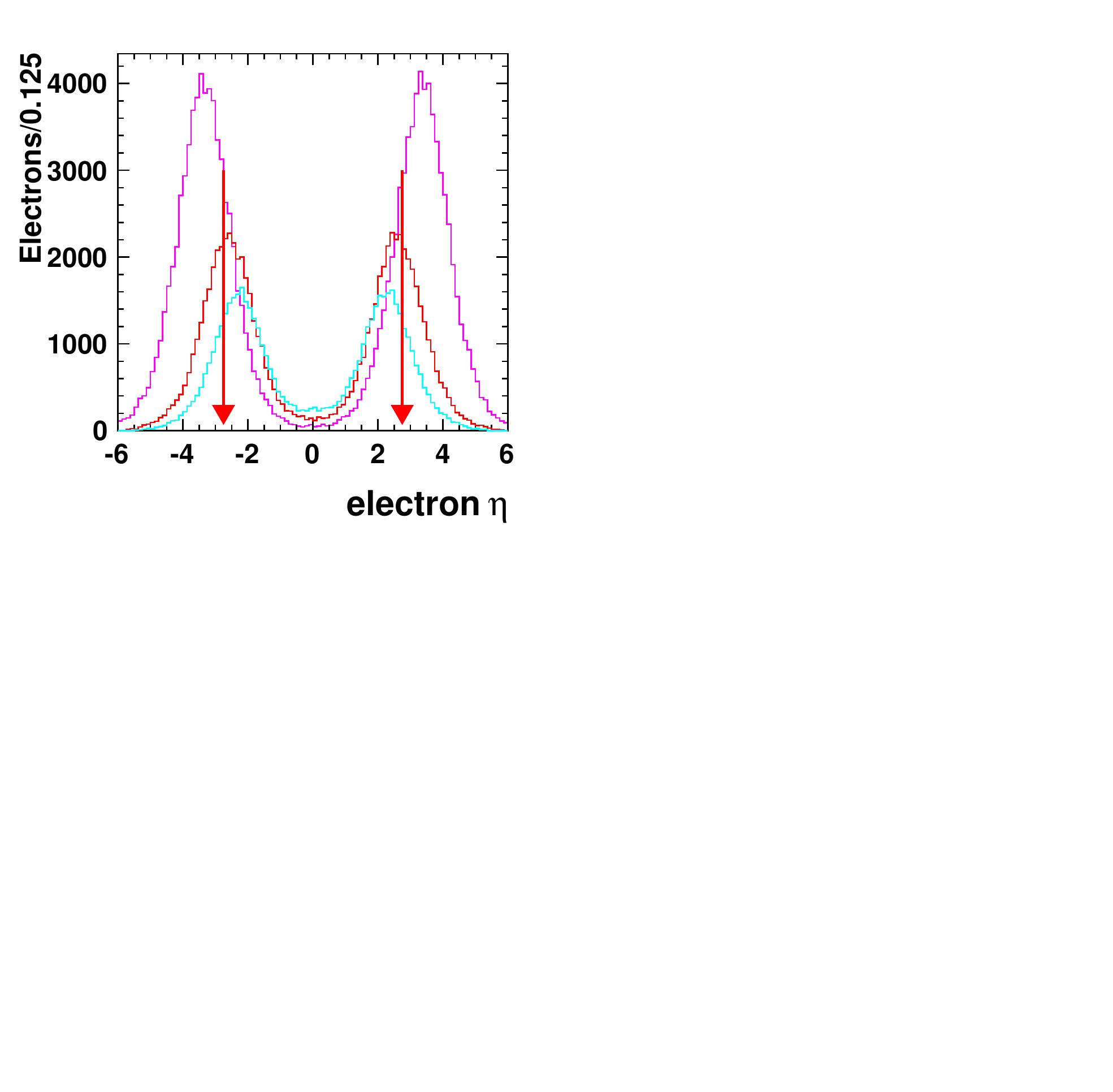}
\begin{textblock}{0.2}(7.1,-5.1)
\bf CLICdp
\end{textblock}
\caption{\label{fig-etael} Scattered beam electrons at the generator level for different centre-of-mass energies (red arrows indicate detector acceptance).}
\end{figure*}

The preselection requires two oppositely-charged electron candidates, separated by $\lvert\Delta\eta\rvert$>1, each with $E$ > 100 GeV.  A further 
requirement is that at least one of the two jets associated with the Higgs decay has a b-tag value greater than 0.65. These cuts lead to signal efficiency of 19.3\%  
resulting in an effective cross-section of 2.6 fb.
The major acceptance loss comes from the geometrical effect of electrons falling outside the detector (Figure \ref{fig-etael}).

The main background process after the preselection is $e^-e^+\rightarrow q\bar{q}l^+l^-$ selected with a 0.2\% efficiency resulting in an effective cross-section of 6.44 fb.

%% file: MVAbb.tex
As a next step in the analysis method, a classification and selection based on a multivariate data analysis is performed. 
A relative likelihood classifier is constructed using four variables that provide separation between signal and background: 

\begin{itemize}
 \item opening distance between the reconstructed electrons: $\Delta R$,
 \item recoil mass of the event determined from the momenta of the reconstructed electrons: $m_{rec}$,
 \item jet transition variable: $-\log y_{34}$,
 \item the invariant mass of the two jets associated with the Higgs decay.
\end{itemize}
  
\begin{figure*}
\centering
\includegraphics[width=5cm, clip, trim = 10cm 12cm 0.5cm 1cm]{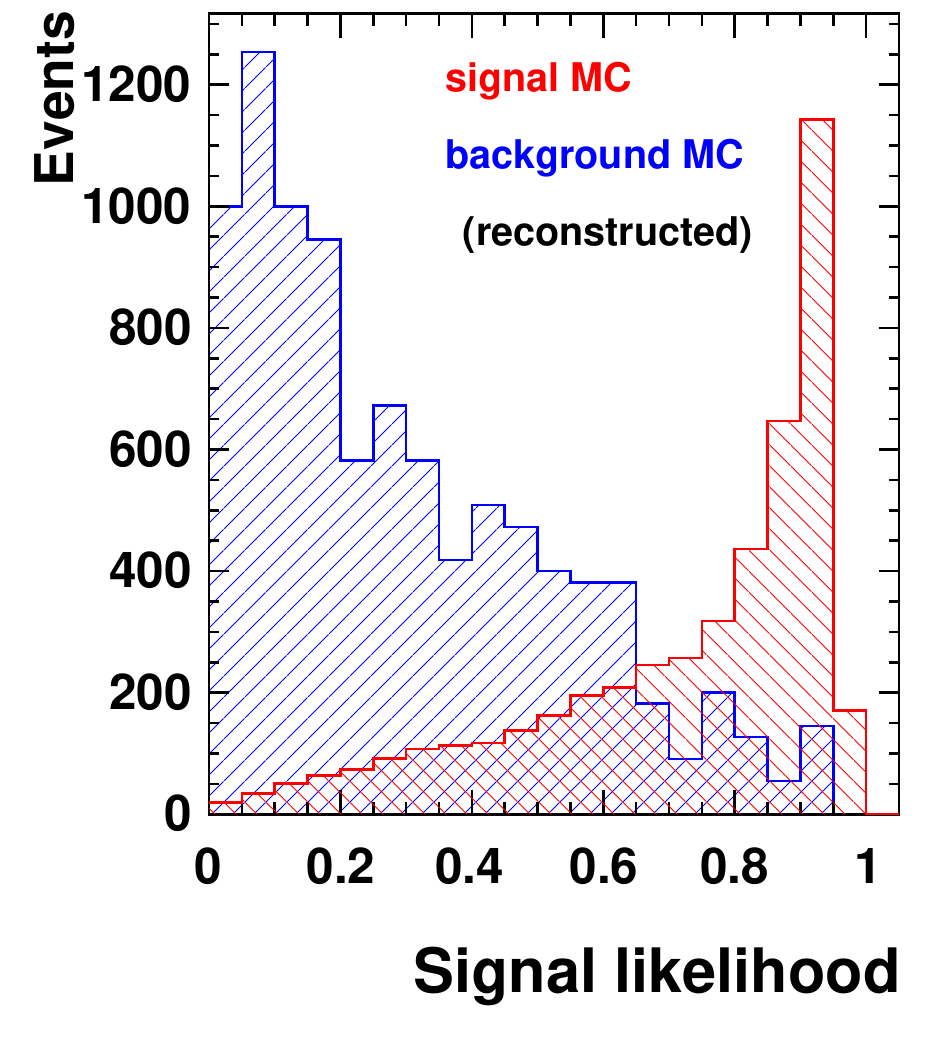}
\begin{textblock}{0.2}(7.5,-3.9)
\bf CLICdp
\end{textblock}
\caption{\label{fig-likelihood} Likelihood distribution for signal (red) and background (blue).}
\end{figure*}

The resulting likelihood distribution is shown in Figure \ref{fig-likelihood} and
gives good separation between signal and background. The result for the statistical uncertainty is reported in Table \ref{table3}.

%% file: results.tex
The results of the measurements of $\sigma(He{^+}e{^-})\times BR(H\rightarrow b\bar{b})$, ${\sigma(H\nu_e\bar{\nu_e})\times BR(H\rightarrow ZZ^\ast \rightarrow q\bar{q}q\bar{q})}$ and
${\sigma(H\nu_e\bar{\nu_e})\times BR(H\rightarrow ZZ^\ast \rightarrow q\bar{q}l^+l^-)}$ are shown in Table \ref{table3}.
All measurements are simulated at 1.4 TeV CLIC collider with unpolarised beams.

The relative statistical uncertainty is 18.3\% and 6\% for the hadronic and semi-leptonic $ZZ^\ast$ decays, respectively.
They are dominated by the limited signal statistics and
the presence of large backgrounds in the $ZZ^\ast \rightarrow q\bar{q}q\bar{q}$ measurement.

The obtained results are included in the the global fit to contribute to the Higgs to $Z$ coupling, $g_{H_{ZZ}}$, and to the total Higgs width ${\ensuremath{\Gamma_{H}}}$.

These results may be improved including tau leptons in the semi-leptonic analysis and trying to further improve the lepton pair efficiency. 

A further improvement may come from the beam polarisation: if 80\% left-handed polarisation of the electron beam is assumed during the entire operation time at 1.4 TeV, the $WW$ 
fusion Higgs production cross-section would be enhanced by a factor 1.8. 

The statistical uncertainty of the measurement $\sigma(He{^+}e{^-})\times BR(H\rightarrow b\bar{b})$ is 1.7\%. This measurement is proportional to 
${g^2_{H_{ZZ^\ast}} \cdot g^2_{H_{b\bar{b}}} / \ensuremath{\Gamma_{H}}}$ and the result is included in global Higgs fit to contribute to the $g_{H_{ZZ}}$ determination.

\begin{table}
\centering
\caption{\label{table3}Summary of the results of the analysis of the ${\sigma(H\nu\bar{\nu})\times BR(H\rightarrow ZZ^\ast)}$ and $\sigma(He{^+}e{^-})\times BR(H\rightarrow b\bar{b})$
measurements at 1.4 TeV CLIC with unpolarised beams. $N_s$ is the number of final signal events, $\epsilon_s$ represent overall signal efficiency of the measurements and 
$\delta(\sigma\times BR)$ is statistical uncertainty on the cross section times branching ratio of the measurements.}
\begin{tabular*}{\columnwidth}{@{\extracolsep{\fill}}llll@{}}
\hline
\multicolumn{1}{c}{}  & \multicolumn{1}{c}{$H\nu\bar{\nu}\rightarrow ZZ^\ast, ZZ^\ast\rightarrow q\bar{q}q\bar{q}$}  & \multicolumn{1}{c}{$H\nu\bar{\nu}\rightarrow ZZ^\ast, ZZ^\ast\rightarrow q\bar{q}l^+l^-$} & \multicolumn{1}{c}{$He{^+}e{^-}\rightarrow b\bar{b}$}  \\
\hline
 $N_s$ & $931 $  & $297 $  & $748 $\\
 $\epsilon_s$   & 18\%  & 33\%  & 19.3\%\\
 $\delta(\sigma\times BR)$ & 18.3\% & 6\%  & 1.7\%\\
\hline
\end{tabular*}
\end{table}